\documentclass[pre,twocolumn,superscriptaddress,showpacs]{revtex4}
 
\usepackage[german,english]{babel}
\usepackage{amssymb}
\usepackage{amsfonts}
\usepackage{amsmath}
 
\begin{document}
 
\title{Tightness of slip-linked polymer chains}

\date{\today}

\author{Ralf Metzler}
\affiliation{Department of Physics, Massachusetts Institute of Technology,
77 Massachusetts Avenue, Cambridge, Massachusetts 02139, USA}
\affiliation{NORDITA, Blegdamsvej 17, DK-2100 K{\o}benhavn {\O}, Denmark}
\author{Andreas Hanke}
\affiliation{Department of Physics, Massachusetts Institute of Technology,
77 Massachusetts Avenue, Cambridge, Massachusetts 02139, USA}
\affiliation{Department of Physics, Theoretical Physics, 1 Keble Road,
Oxford OX1 3NP, United Kingdom}
\author{Paul G. Dommersnes}
\affiliation{Department of Physics, Massachusetts Institute of Technology,
77 Massachusetts Avenue, Cambridge, Massachusetts 02139, USA}
\affiliation{Department of Physics, Norwegian University of Science and
Technology, N-7491 Trondheim, Norway}
\author{Yacov Kantor}
\affiliation{School of Physics and Astronomy, Sackler Faculty of Exact
Sciences, Tel Aviv University, Tel Aviv 69978, Israel}
\affiliation{Department of Physics, Massachusetts Institute of Technology,
77 Massachusetts Avenue, Cambridge, Massachusetts 02139, USA}
\author{Mehran Kardar}
\affiliation{Department of Physics, Massachusetts Institute of Technology,
77 Massachusetts Avenue, Cambridge, Massachusetts 02139, USA}

\begin{abstract}
We study the interplay between entropy and topological constraints for a
polymer chain in which sliding rings ({\em slip-links\/}) enforce pair contacts
between monomers. These slip-links divide a closed ring polymer
into a number of sub-loops which can exchange length between each other.
In the ideal chain limit, we find the joint probability density function for
the sizes of segments within such a slip-linked polymer chain ({\em
paraknot\/}). A particular segment is tight (small in size) or loose (of
the order of the overall size of the paraknot) depending on both the number of
slip-links it incorporates and its competition with other segments.
When self-avoiding interactions are included, scaling arguments can be used to
predict the statistics of segment sizes for certain paraknot configurations.
\end{abstract}

\pacs{05.20.-y, 02.10.Kn, 87.15.-v, 82.35.Lr}

\maketitle

\section{Introduction}

Topological constraints decrease the accessible degrees of freedom of a
polymer chain \cite{flory}. Whether temporary or permanent, they are quite
ubiquitous and effect the typical behaviour of polymers. For instance,
temporary entanglements between chains in a solution or melt of polymers
give rise to reptation dynamics as described by the tube model
\cite{degennes,edwards}. Permanent entanglements, in turn, are central for
the elastic behaviour of rubber (where they are chemically induced during
vulcanisation) \cite{treloar}, gels and Olympic gels \cite{degennes}. Their
influence on the dynamics is reflected by broad relaxation time spectra
\cite{ferry}.

Knots are a particular form of permanent topological entanglements: A
``knotted'' closed chain cannot be reduced to a simple ring
(the so-called unknot) without breaking bonds
\cite{grosberg,reidemeister,kauffman,adams}.
One of the few exact results pertaining to the statistics of knots is that a
sufficiently long closed self-avoiding walk contains knots with probability
one \cite{frisch,sumners}. Thus, topological constraints are inevitably
created during the polymerisation of long closed chains and,
more generally, knots and permanent entanglements are a ubiquitous feature of
multi-chain polymer melts and solutions.

Topological considerations also play a major role in numerous biological and
chemical
systems. For example, the chromosomes forming almost two metres of tangled,
knotted DNA cannot be separated during mitosis, and the genetic code of the
DNA double helix cannot be fully accessed during transcription, in the
presence of knots \cite{alberts,wassermann,wassermann1}. Special enzymes,
namely DNA topoisomerases, are necessary to actively remove knots and
entanglements under consumption of energy from ATP
\cite{alberts,wassermann,wassermann1,strick,yan}.
The interplay between energy and entropy at a fixed topology is relevant to
the secondary structure of RNA which consists of paired segments
interrupted by open loops acting as entropy source \cite{major,bundschuh}. 
Similar issues arise in the
helix-coil transition of DNA \cite{poland,kafri,kafri1,comment}.
Knotted configurations have even been found in some proteins
\cite{proteinbook,taylor}. Dynamically, the presence of knots and their
possible effects on the mobility of biopolymers are essential to the
understanding of their behaviour {\em in vivo\/} or, e.g., as studied by
electrophoresis {\em in vitro\/}
\cite{electrophoresis}. A similar role is played by topological effects for
the translocation of viral and non-viral proteins
\cite{alberts,translocation}, and packaging of DNA \cite{packaging}.
In supramolecular chemistry, molecules with identical bond sequence but
different topology can be produced which exhibit different physical
properties, and mechanically linked molecules open up new vistas in
information processing or nano-engineering \cite{schill,lehn,sauvage}.
Further interest in the theoretical study of the equilibrium behaviour of
polymers with a fixed topology arises from new experimental techniques by
which single molecules can be probed and manipulated
\cite{moerner,micro,afm,tweezer}, providing information on
the mechanical behaviours of knotted and unknotted
biopolymers \cite{arai,rief,raedler}.

Mathematical studies of topological structures date back to Kepler
\cite{kepler}, Euler \cite{euler} and Listing \cite{listing}. Motivated by
Thomson's theory of vertex atoms \cite{kelvin}, systematic studies of knots
were undertaken by Tait, Kirkwood and Little
\cite{tait,kirkman,little,standrews}. Knot theory provides a number of so-called
knot invariants by means of which knots can be classified, such as
the Gauss winding number, the number of essential crossings, or
more refined invariants like knot polynomials \cite{reidemeister,kauffman,adams}.
All permitted configurational changes of a knot can be decomposed
into the three Reidemeister moves \cite{kauffman,adams,reidemeister}.
There exists a fundamental relation between knots and gauge theory as
knot projections and Feynman graphs share the same basic ingredients
corresponding to a Hopf algebra \cite{kauffman}.

Recently there
has been increasing interest in the interplay of topological constraints and
thermal fluctuations; the latter being ubiquitous for dilute or semidilute
polymer solutions or melts at finite temperatures. Statistical
mechanical treatments of permanent entanglements and of knots are, however,
quite difficult since topological restrictions cannot be formulated as a
Hamiltonian problem but appear as hard constraints partitioning the phase
space \cite{degennes,grosberg,vilgisr,kholodenko,REMM}.
Consequently, only a relatively small range of problems have
been treated analytically (see, for instance,
\cite{edwards1,degennes1,vilgis,quake,grosberg1,grosberg2,cates,grone,nechaev}).

To overcome such difficulties in the context of the entropic
elasticity for rubber networks, Ball, Doi, Edwards and coworkers
replaced permanent entanglements by slip-links \cite{edwards2}. Slip-links
enforce contacts between pairs of monomers
but the chain can slide freely through them. Surrogate networks containing slip-links
have been successful in the prediction of important physical quantities of
rubber networks \cite{rubber}. In a similar fashion, we
investigate the statistical behaviour of single polymer chains in which a fixed
topology is created by a number of slip-links. Such {\em ``paraknots''\/} can
be studied analytically using known results for Gaussian random walks in the
ideal chain limit \cite{degennes1,edwards,hughes}. In the language of graphs,
slip-link contacts represent vertices with four outgoing
legs, enabling us to make use of a scaling approach to determine the
leading behaviour in the presence of self-avoiding interactions. The paraknot
approach thus complements our previous study of {\em flat\/} knots in which such
vertices correspond to crossings \cite{slili2d}.

In the following section, we start with a brief summary of conflicting answers
to the question of whether the topological details in a knotted polymer are
localised within a small portion of the chain, and thus segregated from an
unentangled segment. We then introduce the concept of paraknots to study
localisation effects for polymers with fixed topology. Paraknots are first
analysed for ideal chains in Section \ref{section_IIIa}; various contributions
to the joint probability density function (PDF) of segment sizes are easily
separated in this case. There is no similar factorisation of the PDF for
self-avoiding segments, but as discussed in Section \ref{section_IIIb}, scaling
arguments can be used to infer the limiting behaviour of the PDF as one or more
segments contract to small size. The question of the relative sizes of segments
in a paraknot is taken up in Section \ref{tight}. By analysing the
behaviour of the unconditional PDF of a particular segment at small sizes
we can infer whether the segment has the tendency to be tight. Yet, to describe
the actual probability of finding a tight segment, one must consider the
competition between all segments.
For example, even in cases where all segments prefer to be tight
{\em per se}, a given segment can still have a finite probability
of being loose.

\section{From knots to paraknots}
\label{section_II}

\begin{figure}
\unitlength=1cm
\begin{picture}(8,6.2)
\put(-2.2,-20){\includegraphics{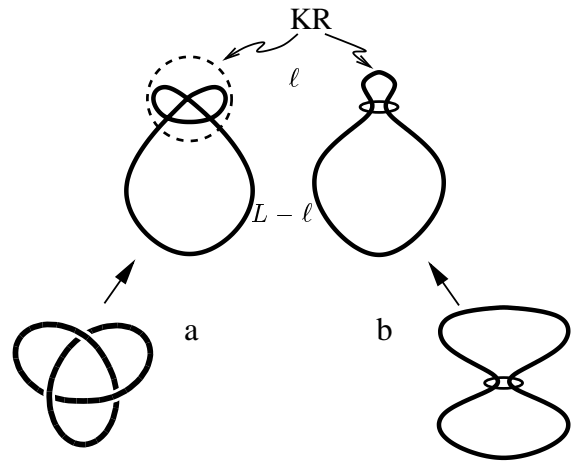}}
\end{picture}
\caption{Depiction of the knot region (KR) for (a) a trefoil knot and (b)
a figure-eight structure in which a slip-link enforces a pair contact
(compare Fig.~\protect\ref{fig1a}). In both cases, a symmetric and
an asymmetric configuration are
shown. The size of the KR is termed $\ell$ and the length of the remaining
simply connected chain is $L-\ell$.
\label{fig0}}
\end{figure}

In Fig.~\ref{fig0}a we depict the projection of a trefoil knot in a symmetric
(bottom) and an asymmetric configuration (top). In the latter
case, we can
introduce a knot region as that part of the knot which contains all
topological details except for the one larger simply connected segment, as
indicated by the dashed line. Initial indications of tight knotted regions are
implicit in 3D Monte Carlo simulations of Janse van Rensburg and Whittington
\cite{janse}, who studied the mean extension of the unknot
and several knot types up to six essential crossings. They found that in
the scaling form $R_g^2\sim(A+BL^{-\Delta})L^{2\nu}$ of the gyration radius both
the prefactor $A$ and the exponent $\nu$ in the leading contribution are {\em
independent\/} of the knot type \cite{janse}. In fact, $\nu$ was found to be
consistent with the known value $\nu=0.588$ of the swelling exponent of a ring
polymer \cite{degennes,grosberg}. (The confluent correction term was estimated 
to decay with $\Delta\approx 1/2$.) Conversely, employing a Flory-type argument
under the assumption that the knot is equally {\em spread out\/} over the
polymer, Quake predicted that the gyration radius should contain the scaling
dependence $R_g\sim\tilde{A}C^{1/3-\nu}L^{\nu}$ on the number of essential
crossings $C$ in the leading order term, i.e., that the amplitude of
$R_g$ decreases with increasing knot complexity \cite{quake}. This result was
supported by his numerical
study of knots up to $8_1$ \cite{quake}, with a different algorithm than used
in Ref.~\cite{janse}. Grosberg {\em et al.} \cite{grosberg2} also make use of a
Flory-type approach assuming that in an evenly delocalised knot the topological
constraints can be replaced by a tube whose radius can be determined from the
aspect ratio of a maximally inflated state. They obtained similar conclusions
to Quake, although they also remark
mhat thermodynamically a segregation into a
simply connected ring polymer and a dense knot region might occur
\cite{grosberg2}.
In a later work, Grosberg states that a more powerful approach is needed to
theoretically decide between the two options \cite{grosberg1}.
More recent numerical studies seem to corroborate the independence of the
gyration radius of the knot type in long enough polymers. Thus, in 3D

\clearpage

\begin{widetext}

\begin{figure}
\unitlength=1cm
\begin{picture}(8,7.68)
\put(-7,-18.4){\includegraphics{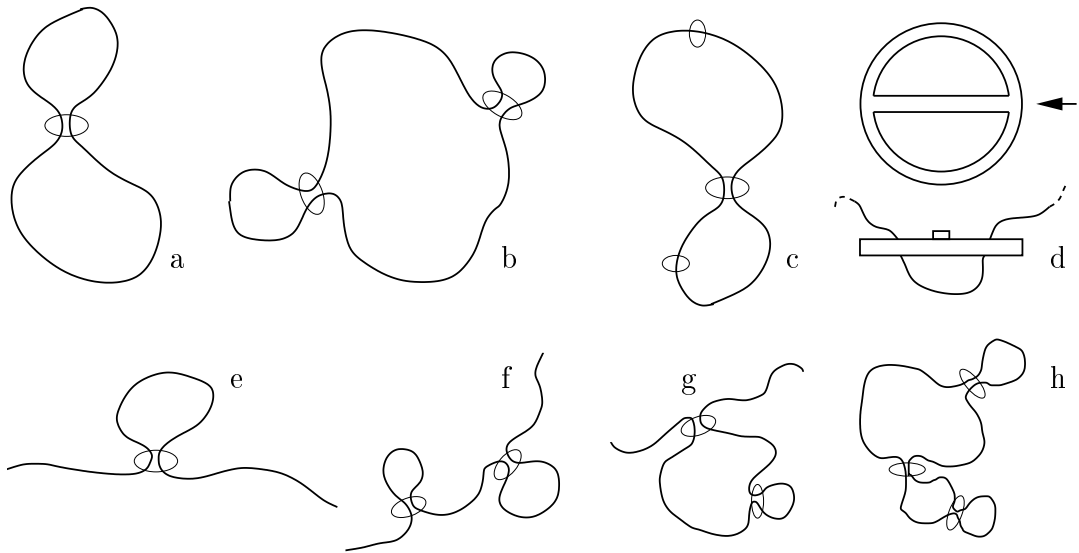}}
\end{picture}
\caption{A collection of different paraknots discussed in the text. (a) The
figure-eight paraknot is formed by placing a slip-link on a ring polymer.
(b) The next higher order paraknot with two slip-links. (c) The figure-eight 
paraknot with two additional sliding rings, one on each separated loop.
(d) Visualisation of a slip-link: the belt buckle shape
allows the chain to slide freely through the slip-link without retracting
entirely. The lower
part corresponds to the view from right as indicated by the arrow. (e) The
lowest order open paraknot. (f) Open paraknot with
two slip-links. (g) Topologically different configuration with two slip-links.
(h) Paraknot necklace with three slip-links.
\label{fig1a}}
\end{figure}

\end{widetext}

\noindent
Orlandini
{\em et al.} calculate in a Monte Carlo study the number of configurations
$\omega_{\cal K}$ of different knot types ${\cal K}$, reporting that $\omega_
{\cal K}\sim\mu^LL^{\alpha-3}$ where both $\mu$ and $\alpha$ are independent of
${\cal K}$ for prime knots, and that an additional factor $L^{n-1}$ occurs
for composite knots with $n$ prime components \cite{orlandini}. These authors
conclude that one or more tight knot regions can move along the perimeter
of a simply connected ring polymer, each prime component being represented
by one knot region \cite{orlandini}. An analogous result
was obtained in 2D by Guitter and Orlandini \cite{orlandini1}. Consistent
with these findings, Katritch {\em et al.} obtained that the knot region is
tight in 3D with a relatively high probability \cite{katritch}. The
investigations of Shimamura and Deguchi \cite{deguchi} corroborate this
picture in obtaining that the gyration radius is independent of the knot type
in some limit of their model.

Why should knots be confined to a small region of the polymer?
Entropic effects give rise to long-range interactions as we demonstrate
for the figure-eight structure sketched in Fig.~\ref{fig0}b in which a
permanent pair contact is enforced by a slip-link, creating two loops of
lengths $\ell$ and $L-\ell$, which can freely exchange length. In the ideal
chain limit, the two loops correspond to returning random walks, i.e.,
the PDF $p(\ell)$ for the size $\ell$ becomes
\cite{degennes,counting}
\begin{equation}
\label{fig8id}
p_{{\rm id}}(\ell)\propto\ell^{-d/2}(L-\ell)^{-d/2},
\end{equation}
where $d$ is the embedding dimension.
The average loop size $\langle\ell\rangle=\int_{a}^{L-a}d\ell\ell
p(\ell)$, where $a$ is a short-distance cutoff set by the lattice constant,
is trivially $\langle\ell\rangle=L/2$, due to the symmetry of the structure.
However, as the PDF is strongly peaked at $\ell=0$ and $\ell=L$, a {\em
typical\/} shape consists of one small ({\em tight\/}) and one large ({\em
loose\/}) loop. For instance, in
$d=3$ the mean size of the smaller loop $\langle\ell\rangle_<$
scales as
\begin{equation}
\label{fig8d3}
\langle\ell\rangle_< \sim a^{1/2} L^{1/2}, \quad d=3,
\end{equation}
which corresponds to {\em weak localisation\/} in the sense that the
smaller loop still grows with $L$, but with an exponent smaller than
one. By comparison, for $d>4$ one encounters $\langle\ell \rangle_<\sim a$,
corresponding to {\em strong localisation} as the size of the smaller
loop does not depend on $L$ but is set by the short-distance cutoff $a$.
On the other hand, for $d = 2$ one finds
$\langle\ell\rangle_< \sim L/|\ln(a/L)|$, such that the smaller loop
is still rather large. However, we will see in the next section that
this is no longer true if we include self-avoiding interactions
for the chains; in that case, the localisation for $d = 2$ is even
stronger than for $d=3$.

Equation (\ref{fig8d3}) shows that the smaller loop, of length $\ell$, of the
figure-eight structure is indeed tight in $d=3$. In fact, for {\em flat\/}
knots rendered as quasi 2D knot projections, it turns out that {\em all\/}
prime knots become tight, and that their leading scaling behaviour corresponds
to the figure-eight structure \cite{slili2d}. This localisation is the
consequence of a delicate interplay between competing effects.
Statistically, the confinement of topological details into
a localised region of the polymer chain is favoured entropically as then the
topological constraints act on a small portion of the chain, exclusively, and
the remaining major part has access to all degrees of freedom of a simply
connected ring polymer. This tendency towards confinement is counteracted by
the internal degrees of freedom of the knot region in which the substructures
can exchange length among each other. It turns out that the tradeoff is in
favour of localisation.

We propose that slip-link structures grasp some essential features of the
statistical behaviour of real knots and therefore call them {\em paraknots.}
Some elementary examples of paraknot structures which will be discussed in
the following Sections are displayed in Fig.~\ref{fig1a}. The slip-links in
these configurations can be viewed as little rings which enforce pair contacts
within the chain such that the loop formed by the slip-link is not allowed
to fully retract. In a simulation, this latter property can be included by
a belt buckle shape as sketched in Fig.~\ref{fig1a}d. In a paraknot, one or
several loops may be cut, creating open chain segments as in Figures
\ref{fig1a}e to g. Such ``open'' paraknot
types can be stabilised (i.e., an open end prevented from escaping
through a slip-link) by attaching ``stoppers'' to the open ends, such as
latex microspheres, ring-molecules, or C$_{60}$ balls, as known from
supramolecular chemistry \cite{lehn,sauvage}.

Paraknots are tractable exactly in the ideal chain limit, and by
scaling theories in the self-avoiding domain. In the following, we investigate
the statistical description and the localisation properties of several paraknot
structures.

\section{Statistical weights of paraknot structures}

\begin{figure}
\unitlength=1cm
\begin{picture}(8,12.4)
\put(-2.4,-13.8){\includegraphics{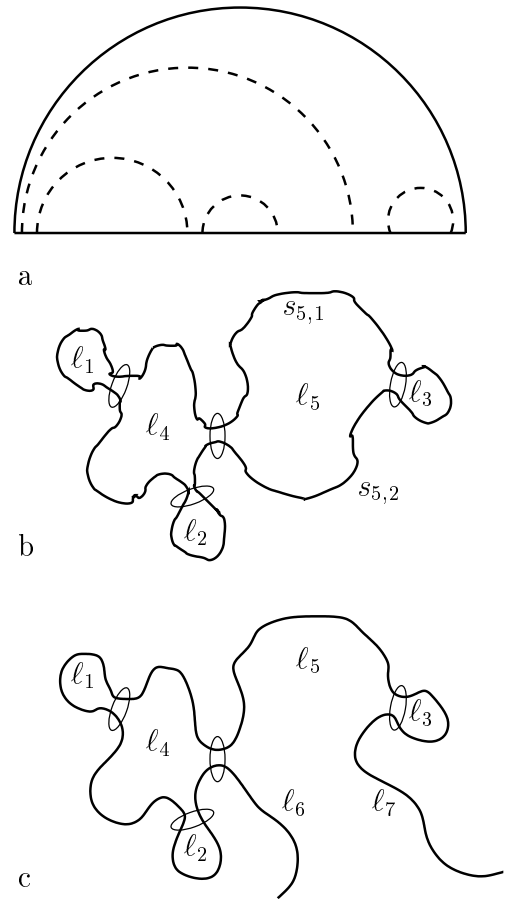}}
\end{picture}
\caption{(a) Arc diagram for the construction of a closed paraknot from a
polymer ring (full line). The dotted lines indicate which points of the
chain are connected to each other by slip-links. (b) The paraknot resulting
from this procedure. (c) Open paraknot obtained by cutting loop 5, creating
two open legs. Note that individual connectors (dashed lines) are not
allowed to intersect each other, i.e., the paraknot contains
unconcatenated loops.
\label{fig1}}
\end{figure}

A general paraknot can be constructed, as shown in Fig.~\ref{fig1}, from an
arc diagram similar to those used to classify the secondary structure of RNA
\cite{major,bundschuh}. Such an arc diagram is the blueprint of the associated
paraknot, and it features the original loop into which slip-links are
introduced by connecting pairs of monomers through the dashed lines. To
simplify the analysis, we only consider paraknots with unconcatenated loops,
i.e., the arcs are not allowed
to intersect each other. In the RNA language, this means that pseudoknots are
not permitted \cite{major,bundschuh}. With this restriction, the joint PDF for
the sizes of various segments simplifies to a product of contributions from
loops, in the case of ideal (Gaussian) polymers. As discussed next, in the case
of self-avoiding walks, only scaling information is available in the limit when
segments contract to small sizes.

\subsection{Ideal chains}
\label{section_IIIa}

For ideal chains, analytical calculations are rather straightforward for
non-crossing arc diagrams (similar to Hartree graphs).
For instance, consider the paraknot
shown in Fig.~\ref{fig1} for fixed loop lengths
$\ell_1,\ldots,\ell_5$. The key observation is that for ideal
chains the degrees of freedom associated with the individual loops
are decoupled from one another, so that the PDF of the paraknot factorises
into the corresponding loop contributions. Following the above
example, a general paraknot ${\cal P}$ can be described by the set
$\{\ell_1,\ell_2,\ldots,\ell_m\}$ of individual loop lengths (also including
end-to-end lengths in case of linear
segments) under the constraint $L=\sum_i\ell_i$,
where the contributions from the individual loops (or linear segments)
factorise. In equilibrium, these lengths are thus distributed
according to the joint PDF
\begin{equation}
\label{pdf}
p_{{\cal P}}(\ell_1,\ell_2,\ldots,\ell_m)\propto\delta\left(L-\sum_{i=1}^{m}
\ell_i\right)\prod_{i=1}^{m}\ell_i^{-\theta_i} \, ,
\end{equation}
where the exponents $\theta_i$ are constructed from the following
contributions \cite{pdfrem}:

{\bf (i) Connectivity factor:} This factor accounts for the configurational
entropy of a given loop (or linear segment) of length $\ell$. For a loop, the
connectivity factor follows from the return probability of a
Gaussian random walk, which is $\sim \ell^{-d/2}$. The absence of any
constraint for a linear segment corresponds to a factor $\sim\ell^0$.

{\bf (ii) Sliding entropy:} A given loop (or linear segment) of length $\ell$
has additional degrees of freedom associated with the slip-links which slide
on it. This is due to the relative motion of these slip-links along the segment.
The presence of $n$ slip-links on a loop (or linear segment) thus leads
to a factor of $\ell^{n-1}/(n-1)!$ (or $\ell^n/n!$).
Additional degrees of freedom in the form of sliding rings confined to a
given segment as depicted in Fig.~\ref{fig1a}c enter the PDF analogously.

{\bf (iii) Energetic factors:} If an external force is applied to the paraknot,
a Boltzmann weight enters the expression for the size distribution. For
instance, if an open paraknot is pulled with a constant force ${\bf f}$, this
weight corresponds to the average $\overline{\exp\{\beta{\bf f}\cdot{\bf r}\}}=
\exp\{\beta^2f^2\overline{r^2}/(2d)\}$ where the overline indicates the
average over all end-to-end distances ${\bf r}$ of the backbone segment, and
$\beta\equiv1/(k_BT)$. (Such effects will be relegated to a future work
\cite{pull}.)

\subsection{Self-avoiding chains}
\label{section_IIIb}

If self-avoiding constraints for the chains are included, the above reasoning
for ideal chains, in particular the factorisation of the PDF,
is no longer valid in general since now every loop or segment
of the paraknot interacts with all the others. However, progress can be made,
allowing for quantitative predictions of the leading scaling behaviour of a
given paraknot, by employing the
scaling theory for self-avoiding polymer networks developed by Duplantier
\cite{duplantier},
Sch{\"a}fer {\em et al.} \cite{schaefer}, and Ohno and Binder \cite{binder}.
This approach has recently been applied to the study of DNA denaturalisation
by Kafri {\em et al.} \cite{kafri,kafri1} and to the study of 2D knots
\cite{slili2d}.

\begin{figure}
\unitlength=1cm
\begin{picture}(8,4.6)
\put(-2.4,-21.5){\includegraphics{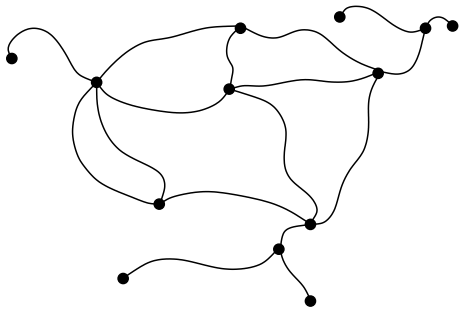}}
\end{picture}
\caption{Polymer network ${\cal G}$ with vertices ($\bullet$) of different
order ($n_1=5$, $n_3=4$, $n_4=3$, $n_5=1$).
\label{netw}}
\end{figure}

A general polymer network ${\cal G}$ like the one depicted in Fig.~\ref{netw}
consists of a number of vertices
which are joined by ${\cal N}$ chain segments of individual lengths
$s_1,\ldots,s_{\cal N}$ whose total length is $L=\sum_{i=1}^{\cal N} s_i$.
The number of configurations of such a network scales as
\cite{duplantier,schaefer,binder}
\begin{equation}
\label{network}
\omega_{\cal G}\sim \mu^Ls_{\cal N}^{\gamma_{\cal G}-1}{\cal Y}_{\cal G}
\left(\frac{s_1}{s_{\cal N}},\ldots,\frac{s_{{\cal N}-1}}{s_{\cal N}}\right),
\end{equation}
where $\mu$ is the effective connectivity constant for self-avoiding walks
and ${\cal Y}_{\cal G}$ is a scaling function. The topology of
the network is reflected in the exponent
\begin{equation}
\gamma_{\cal G}=1-d\nu{\cal L}+\sum_{N\ge 1}n_N\sigma_N,
\end{equation}
where ${\cal L}=\sum_{N\ge 1}(N-2)n_N/2+1$ is the number of
independent loops, $n_N$ is the number of vertices with $N$ legs,
and $\sigma_N$ is an exponent connected to an $N$-vertex.
The PDF of the paraknot then follows from the number of configurations
$\omega_{\cal G}$ by normalisation with respect to the variable segment lengths.

{\bf (i) Connectivity factor:} To illustrate how the connectivity factors
(of the form $\sim\ell^{-d/2}$ and $\sim\ell^0$ for closed loops and linear
segments in the Gaussian case) are modified, let us consider the cases
of the figure-eight paraknot (Fig.~\ref{fig1a}a), and its open counterpart
(Fig.~\ref{fig1a}e).

The figure-eight paraknot corresponds to a network with two loops of lengths
$\ell$ and $L-\ell$, respectively, and one vertex with 4 legs. We thus obtain
\begin{equation}
\label{y8}
\omega_8 \sim \mu^L(L-\ell)^{-2d\nu+\sigma_4}{\cal Y}_{8}
\left(\frac{\ell}{L-\ell}\right),
\end{equation}
for the configuration number. Now we use the {\em a priori\/} assumption that
$L-\ell\gg\ell$. Then, the large loop should behave like a ring polymer of
length $L-\ell$, i.e., it should
contribute to $\omega_{\cal G}$ in the scaling form $(L-\ell)^{-d\nu}$
\cite{kafri}. This can only be fulfilled if the scaling function behaves like
${\cal Y}_{8}(x)\sim x^{-c}$ with $c=d\nu-\sigma_4$. The final result for
the number of configuration of the self-avoiding figure-eight paraknot then
becomes
\begin{equation}
\label{acht}
\omega_8 \sim \mu^L(L-\ell)^{-d\nu}\ell^{-d\nu+\sigma_4}.
\end{equation}
In $d=2$, with $\nu=3/4$ and $\sigma_4=-19/16$ \cite{duplantier,schaefer,kafri},
we therefore find that the small loop scales like $\ell^{-c}$ where
$c=2.6875$. In $d=3$, we obtain the exponent $c\approx 2.24$ using
$\nu=0.588$ and $\sigma_4\approx-0.46$
\cite{duplantier,schaefer,kafri,nienhuis}. The strong localisation which
obtains for both $d=2$ and $d=3$ is the {\em a posteriori\/} justification of
the $L-\ell\gg\ell$ assumption, and the procedure is therefore
self-consistent. Note that in the presence of self-avoiding
constraints, the localisation is stronger in 2D than in 3D,
in contrast to the ideal chain case (see Sec. \ref{section_II}).

\begin{widetext}
 
\begin{figure}
\unitlength=1cm
\begin{picture}(18,7.72)
\put(-2,-18.32){\includegraphics{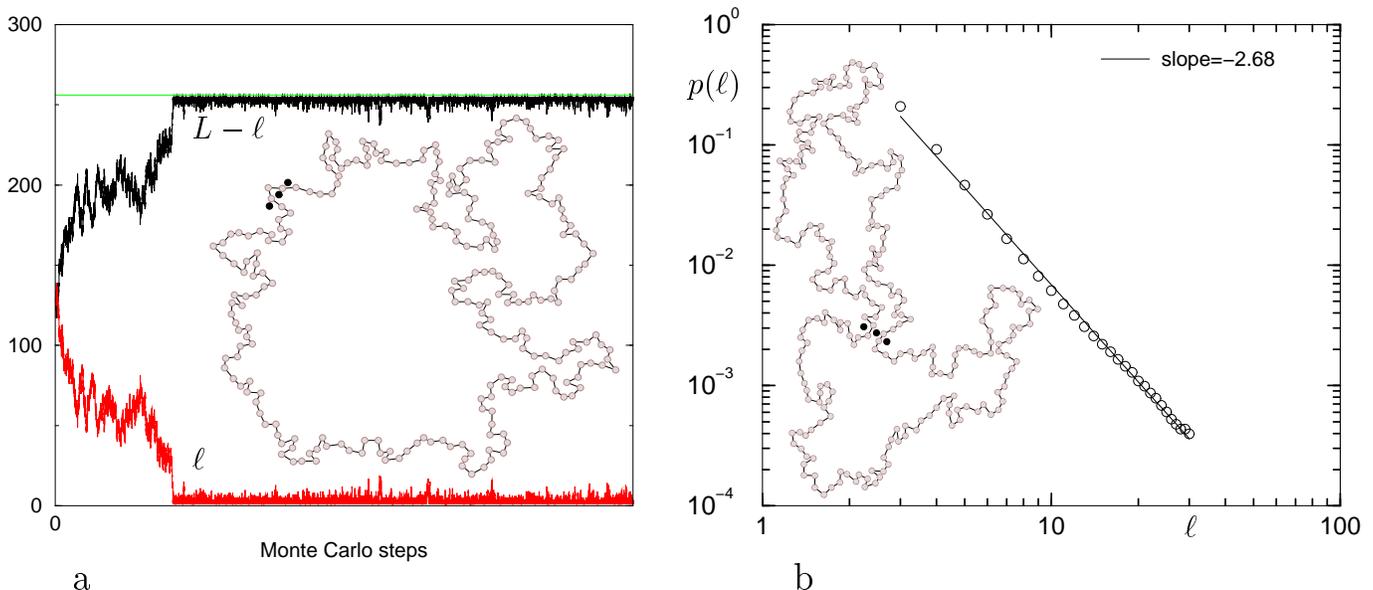}}
\end{picture}
\caption{Monte Carlo simulation of a figure-eight paraknot in 2D. (a) Loop
sizes $\ell$ and $L-\ell$ as a function of Monte Carlo steps for 256 monomers.
In the inset, a typical equilibrium configuration is shown. The slip-link is
made up of the three tethered beads rendered black which constitute the 2D
version of the belt buckle shape depicted in Fig.~\protect\ref{fig1a}d.
(b) Probability density
for the size $\ell$ of the smaller loop for a figure-eight with 512 monomers.
The inset shows an intermediate configuration reminiscent of the symmetric
initial condition.
\label{simu}}
\end{figure}
 
\end{widetext}

We performed a Monte Carlo analysis of the elementary slip-link in 2D with
a standard bead-and-tether chain.
In Fig.~\ref{simu}a, we show the equilibration of a symmetric
initial configuration and its fluctuations as a function of Monte Carlo steps.
Clearly, the separation into two length scales is fast and fluctuations are
relatively small. The size distribution of the small loop is displayed in
Fig.~\ref{simu}b. From the plot, we realise that the scaling be-

\clearpage
 
\noindent
haviour is surprisingly well fulfilled,
and that the predicted value is reproduced in good agreement.
This result was corroborated experimentally for a figure-eight necklace chain
on a vibrating plate \cite{bennaim}.

Compare this finding to the lowest order open paraknot (o) sketched in
Fig.~\ref{fig1a}e. Apart from the vertex with 4 legs, there are
two vertices with one leg, one for each of the two ends of the linear
chain segment, thus yielding
\begin{equation}
\omega_{o} \sim\mu^L(L-\ell)^{1-d\nu+2\sigma_1+\sigma_4}{\cal Y}_{o}
\left(\frac{\ell}{L-\ell}\right).
\end{equation}
We again assume {\em a priori\/} that the open chain segment is the overall
dominating structure of size $L-\ell$. It thus enters into $\omega_{o}$ in
the form $(L-\ell)^{\gamma}$ where $\gamma=2\sigma_1+1$ is the configuration
exponent \cite{degennes1,descloiseaux,duplantier}, which implies
\begin{equation}
\omega_{o} \sim \mu^L(L-\ell)^{\gamma}\ell^{-d\nu+\sigma_4}.
\end{equation}
Thus, we find the same loop closure exponent as for the figure-eight
structure above [see Eq.~(\ref{acht})].
This is not surprising, as the small loop is statistically
independent of the large structure; in other words, the topological
exponent $\sigma_4$ stems from the nature of the vertex, which is a
{\em local\/} quantity.

The closure factor for a loop in such simple geometries as the
figure-eight or the lowest order open paraknot is therefore given by
$\sim\ell^{-d\nu+\sigma_4}$; the factor for the degrees of freedom
of the linear chain segment in the latter enters as $\sim(L-\ell)^{\gamma}$.

{\bf (ii) Sliding entropy:} Consider the paraknot shown in Fig.~\ref{fig1}b.
It constitutes a polymer network ${\cal G}$ in which four 4-vertices
(corresponding to the slip-links) are joined by a number of chain segments
$s_i$ with $\ell_1=s_1$, $\ell_5=s_{5,1}+s_{5,2}$, etc. 
Since the loops are non-concatenated, it is possible to integrate
the right hand side of Eq.~(\ref{network}) for {\em fixed\/} loop
lengths $\ell_1,\ldots,\ell_5$ over some of the segment lengths
$s_i$ in such a way that the resulting expression depends
on the loop lengths only, i.e.,
\begin{equation}
\label{loops}
\omega_{\cal G}^{(\ell)} \sim \mu^L \ell_1^{\gamma_{\cal G}-1}
\ell_4^2\ell_5{\cal X}_{\cal G}
\left(\frac{\ell_1}{\ell_5},\ldots,\frac{\ell_4}{\ell_5}\right),
\end{equation}
where the scaling function ${\cal X}_{\cal G}$ depends on
ratios of loop lengths
(again, this procedure would not be possible if the paraknot contained
concatenated loops). The superscript $(\ell)$ on $\omega_{\cal G}$
indicates that here the loop lengths $\ell_1,\ldots,\ell_5$ are
fixed. The factors of $\ell_4^2$ and $\ell_5$ in the
above expression correspond to the
sliding entropy already encountered for ideal chains, see (ii)
in Sec. \ref{section_IIIa}.

\section{Tight or loose}
\label{tight}

When is a certain segment of such a paraknot network tight? {\em A priori,}
this can be investigated by integration of the joint PDF over all other
segments. The
result depends on both the local property of the segment itself, i.e.,
on its exponent in the joint size distribution, and its global interplay with
other segments in the paraknot in their cooperative search for the entropically
favoured configuration. In practice, the unconditional PDF can only be obtained
for Gaussian paraknots in which the joint PDF has the multiplicative form in
Eq.~(\ref{pdf}). Such calculations are not possible for general self-avoiding
paraknots, as the unconditional PDF comes out from the specific scaling
analysis for a given
paraknot configuration. We therefore address phantom and self-avoiding
cases separately.

\subsection{Ideal chains}

Consider the joint PDF in Eq.~(\ref{pdf}) for a given paraknot ${\cal P}$
with ${\cal N}$ segments. According to the previous section, the exponents
$\theta_i$ in Eq.~(\ref{pdf}) are given by $\theta_i=d/2-(n_i-1)$ if $\ell_i$
is a loop or by $\theta_i=-n_i$ if it is a linear segment, where $n_i$ is
the number of slip-links and sliding-rings connected to this segment. From
Eq.~(\ref{pdf}), we consider a segment to be loose {\em per se\/} if its
exponent
$\theta_i\le 1$, otherwise it is tight (and ``super-tight'' if $\theta_i>2$).
We can now distinguish between three different global situations:

{\bf (i)} There is one loose segment and all others are tight. This case occurs
if in ${\cal P}$ only one segment with $\theta_{\cal N}\le 1$ exists, while all
others have $\theta_i>1$.

{\bf (ii)} There is more than one loose segment and possibly some tight
segments. In this case, the loose segments compete for the length $L$. On the
average, if there are ${\cal M}$ loose segments, the characteristic length of
any specific segment is $\langle\ell_j\rangle=\frac{1-\theta_j}{\sum_{i=1}^
{\cal M}(1-\theta_i)}L$, which is always larger than 0 and smaller than $L$.
The ratio of characteristic lengths for a pair of segments $j,k$ is then given
by $\langle\ell_j\rangle:\langle\ell_k\rangle=(1-\theta_j):(1-\theta_k)$
\cite{rem}.

{\bf (iii)} All segments are tight {\em per se\/} in the sense that all
$\theta_i>1$. In this case, a symmetry-breaking occurs and one segment becomes
large. The unconditional PDF for each segment will have two peaks corresponding
to tight or loose configurations.

In a paraknot which  contains one or multiple open segments, the open
segments are always loose and therefore only cases (i) and
(ii) can arise: Depending on the exponents of the closed
loops in such a structure, these loops may either be loose or tight. Note that
cases (i) and (iii) exhibit one large loop, in (i) this is the loose segment
and in (iii) it is the one segment which becomes large by symmetry breaking.
As all other segments of paraknots which belong to these classes
are tight, the gyration radius of such paraknots is, to leading order, the
same as for an unknot of length $L$. For paraknots belonging to class (ii),
segments of comparable size make up the gyration radius. Depending on the
details of the structure, the gyration radius should be given by similar
expressions to those developed in Refs. \cite{quake,grosberg}. Thus, the
gyration radius decreases with increasing number of loose segments.

\subsection{Self-avoiding chains}

As mentioned, generalising the previous classification to self-avoiding
structures is not straightforward. Let us therefore consider the tightness
of segments in a self-avoiding paraknot by means of three examples.

\subsubsection{The Round-Table Configuration}

\begin{figure}
\unitlength=1cm
\begin{picture}(8,7.4)
\put(-2.4,-18.8){\includegraphics{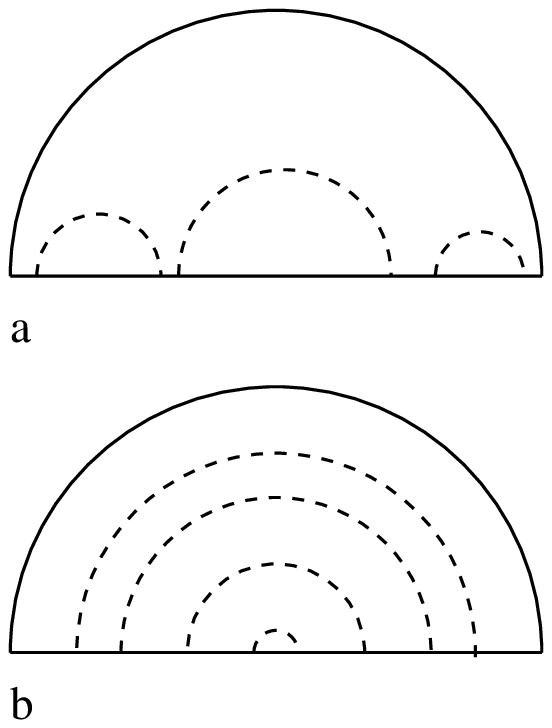}}
\end{picture}
\caption{Arc diagram for (a) the Round Table configuration with $n=3$ fringe
loops, and (b) the Necklace paraknot.
\label{rtarc}}
\end{figure}
This configuration corresponds to arc diagrams in which none of the connecting
arcs is located inside another arc, as shown in Fig.~\ref{rtarc}a. The resulting
paraknot features a number $n$ of loops located at the fringe of a central
loop, as depicted to lowest order in Fig.~\ref{fig1a}b. As the loops are
independent and are connected to one slip-link each, they enter the joint PDF
of loop sizes through the loop closure factor $\sim\ell_i^{-d\nu+\sigma_4}$ as
was found for the figure-eight paraknot. They are therefore super-tight for
self-avoiding chains. Conversely, the central loop has access to additional
degrees of freedom stemming from the relative motion of the slip-links along
its circumference such that its exponent becomes $\theta=d\nu-(n-1)$. Note
that in 2D, the Round Table configuration corresponds to the leading order
behaviour of a composite knot. Each fringe loop, that is, corresponds to a
prime component. The additional degrees of freedom coming about by the
relative motion of the fringe loops in this configuration correspond to the
enhancement of accessible numbers of configurations for knotted chains as
measured by Orlandini {\em et al.} \cite{orlandini,orlandini1}.

\subsubsection{The figure-eight cactus with attached loops}

Consider a figure-eight paraknot in which $n$ and $m$ additional (external)
loops are
attached to the two loops (compare Fig.~\ref{fig1}b in which $n=2$ and $m=1$).
The external loops are all strongly localised and give rise to additional
sliding entropy for the figure-eight loops;
by choosing different values for $n$ and $m$, one obtains various
localisation properties for the figure-eight loops.
For instance, for $n=2$ and $m=1$, $\ell_4$ is loose and $\ell_5$ is weakly
localised; for $n,m\ge 2$, both $\ell_4$ and $\ell_5$ are loose, i.e.,
proportional to $L$, so that the gyration radius of the paraknot is
smaller than for a simple ring of same length $L$. In this case,
the joint PDF does not factorise, but the scaling function ${\cal Y}_{8}$
from Eq.~(\ref{y8}) enters. Depending on the
details of the structure, the gyration radius should be given by similar
expressions as developed in Refs. \cite{quake,grosberg}. Note that analogous
loosening of loops can be achieved for open paraknots of the type sketched
in Fig.~\ref{fig1a}g.

\subsubsection{The Necklace}

Finally, let us explore the Necklace structure from Fig.~\ref{fig1a}h
whose corresponding arc diagram is shown in Fig.~\ref{rtarc}b. In this
configuration, the two end-loops are strongly localised;
the other $m=n-2$ inner loops have two neighbours. By necessity,
one of the inner loops has to be loose (size $L$), and
has sliding entropy with weight $\sim L^{1-d\nu}$. On each side
of the large loop there is a number of other inner loops,
arranged in a hierarchy of shapes of the
type .oOoo. (. = strongly localised end-loop, o = weakly localised loop, O =
large inner loop), statistically changing to .oooO. {\em etc.}
If one focuses on one particular
inner loop, there is a $1/m$ chance to find this loop large
with the (integrated) PDF $\sim L^{1-d\nu}$.
Note that a complete analysis of this relatively simple and symmetric
structure is already quite nontrivial.

The Necklace structure can be closed by an additional slip-link connecting the
two outer loops. This forms a symmetric configuration in which all loops are
{\em a priori\/} equivalent. This paraknot is equal to the network
studied for flat prime knots in Ref. \cite{slili2d}. Accordingly, the closed
Necklace structure is, to leading order, contracted to the figure-eight
paraknot in both 2D and 3D.

\section{Conclusions}

We have presented a systematic study of slip-link structures which we
call paraknots. Paraknots are relatively easy to deal with analytically
and may provide information on the generic
interplay between entropy and fixed topology in polymer chains and networks.

Paraknots composed of ideal chains are described by Gaussian propagators for
which calculations are reasonably straightforward. Simple paraknots in 2D and
3D are only marginally or weakly localised whereas localisation is strong in
higher dimensions. More complicated paraknots in which individual loops are
connected to more than one slip-link show less localisation due to the
additional degrees of freedom brought about by the relative motion of the
slip-links or additional sliding-rings on a loop. If self-avoiding effects are
considered, simple paraknots are strongly localised even in 2D and 3D. The
scaling exponents involved can be
obtained from Duplantier's theory for general polymer networks. For the
figure-eight paraknot, we have confirmed the scaling exponent through Monte
Carlo simulations. Localisation in self-avoiding paraknots becomes weakened if
more than one substructure has additional degrees of freedom, in analogy to the
ideal chain case. This observation pertains to arbitrary
topological polymer networks.

The tightness of paraknots in 2D quantifies the
strong localisation for flat knots observed by Guitter and Orlandini
\cite{orlandini1}. Whereas we cannot infer definitive statements on
3D knots from our analysis, the correspondence between figure-eight paraknot
and the leading order behaviour of prime knots (and between the Round
Table configuration and composite knots) in 2D suggests that similar
tightness could be observed in 3D as well. This is consistent with the
findings of Janse van Rensburg and Whittington \cite{janse}, Orlandini {\em et
al.} \cite{orlandini} and Katritch {\em et al.} \cite{katritch}, and it
differs from the conclusions of Quake \cite{quake}.

Additional energetic effects due to bending and the presence of (screened)
electric charges are relevant for many systems, especially in biology. In
so far as these effects can be accommodated by the introduction of a
persistence length, they should not affect our results in the long-chain
limit. However, they determine the crossover size for the onset of the
long-chain limit in the polymer.
For the particular case of the 2D trefoil knot the results of our previous
analysis suggest that the continuum limit is reached for chains with 512
monomers, whereas for the figure-eight paraknot even chain lengths of 128
seem to be sufficient. Thus, in a DNA double helix for which the persistence
length is of the order of 100 base pairs (bp), one may expect to see
localisation behaviours in {\em simple\/} topologically entangled states
for strand lengths of the order of 10 to 50kbp, corresponding to a length of
5 to 25$\mu$m \cite{siggia}. For shorter DNA strands, it is to be
expected that finite size effects prevail, and thus the knots or other
topological details will be spread out over a considerably larger part of the
entire chain.

\begin{acknowledgments}
We acknowledge financial support from the National Science Foundation
(DMR-01-18213 and PHY99-07949) and the US-Israel Binational Science
Foundation (BSF) grant No.~1999-007. AH and RM acknowledge financial
support from the Deutsche Forschungsgemeinschaft (DFG). AH also
acknowledges financial support by the Engineering and
Physical Sciences Research Council through grant GR/J78327.
PGD acknowledges financial support from the Research Council of Norway.
\end{acknowledgments}


\begin{thebibliography}{99}
\bibitem{REMM} For comparison, self-avoidance in 3D is usually treated as a
perturbation, i.e., as a ``soft constraint'', in analytical studies
\protect\cite{degennes}.
\bibitem{counting} Here and in the following we consider two configurations
of a polymer chain different if they cannot be matched by translation. In
addition, the origin
of a given structure is fixed by a vertex point, i.e., a point where
several legs of the polymer chain are joint. For a simply connected
ring polymer, such a vertex is a two-vertex anywhere along the chain.
\bibitem{pdfrem}
Note that the above factorisation does not take place in structures where
additional links are introduced, such as connecting the two corner loops
in the necklace configuration sketched in Fig.~\ref{fig1a}h;
this creates conditional probabilities, and the associated arc
diagrams no longer correspond to Hartree graphs.
\bibitem{rem} If the segments $\ell_1,\ell_2,\ldots,\ell_{\cal M}$ are loose,
the cutoff $a$ is no longer relevant and characteristic lengths can be
calculated by multiple integrals of the form
\begin{eqnarray*}
I_j(L)=\int_0^Ld\ell_1\ell_1^{-\theta_1}\ldots
\int_0^{L-\sum_{i=1}^{j-1}\ell_i}d\ell_j\ell_j^{1-\theta_j}
\ldots\hspace*{0.5cm}\\
\times\int_0
^{L-\sum_{i=1}^{{\cal M}-2}\ell_i}d\ell_{{\cal M}-1}\ell_{{\cal M}-1}^{
-\theta_{{\cal M}
-1}}\left(L-\sum_{i=1}^{{\cal M}-1}\ell_i\right)^{-\theta_{{\cal M}}},
\end{eqnarray*}
whose Laplace transform ($L\to u$)
\[
\widetilde{I_j(u)}= u^{\sum_{i=1}^{{\cal M}}(\theta_i-1)-1}
\Gamma(2-\theta_j)\prod_{i\neq j}\Gamma(1-\theta_i)
\]
yields in a simple product form by the integration theorem for the first
integral and the convolution theorem for the remaining factors, and
therefore
\[
I_j(L)=L^{\sum_{i=1}^{{\cal M}}(1-\theta_i)}\frac{\Gamma(2-\theta_j)
\prod_{i\neq j}\Gamma(1-\theta_i)}{\Gamma(1+\sum_{i=1}^{{\cal M}}
[1-\theta_i])}.
\]
With similar expressions for the normalisation, one obtains the results quoted
in the text.
\end{thebibliography}
\end{document}